\documentclass[%
 reprint,
 amsmath,amssymb,
]{revtex4-2}
\usepackage{graphicx}
\usepackage{dcolumn}
\usepackage{bm}
\usepackage[mathlines]{lineno}

\usepackage[utf8]{inputenc}
\usepackage[T1]{fontenc}
\usepackage{mathptmx}
\usepackage{etoolbox}
\usepackage{xcolor}
\usepackage{natbib}
\usepackage{comment}

    \newcommand{\vct}[1]{\boldsymbol{#1}}
\makeatletter
\def\@email#1#2{%
 \endgroup
 \patchcmd{\titleblock@produce}
  {\frontmatter@RRAPformat}
  {\frontmatter@RRAPformat{\produce@RRAP{*#1\href{mailto:#2}{#2}}}\frontmatter@RRAPformat}
  {}{}
}%
\makeatother
\begin{document}

\preprint{APS/123-QED}

\title{\textbf{Kinetic Scale Energy Budget in Turbulent Plasmas: Role of Electron to Ion Temperature Ratio} 
}%

\author{Subash Adhikari$^*$}
\email{Corresponding author: subash@udel.edu}
\affiliation{Department of Physics and Astronomy, University of Delaware, Newark, DE 19711}
\author{M. Hasan Barbhuiya}%
\affiliation{ Department of Physics and Astronomy, Clemson University, Clemson, SC 29634}%

\date{\today}

\begin{abstract}
The dissipation mechanisms in weakly collisional plasmas have been a longstanding topic of investigation, where significant progress has been made in recent years. 
A recent promising development is the use of the ``scale-filtered'' Vlasov–Maxwell equations to fully quantify the scale-by-scale energy balance, a feature that was absent when using fluid models in kinetic plasmas. 
In particular, this method reveals that the energy transfer in kinetic scales is fully accounted for by the scale-filtered pressure–strain interaction. Despite this progress, the influence of ion–electron thermal disequilibrium on the kinetic-scale energy budget remains poorly understood. Using two-dimensional fully kinetic particle-in-cell simulations of decaying plasma turbulence, we systematically investigate the pressure–strain interaction and its components at sub-ion scales by varying electron-to-ion temperature ratios. Our analysis focuses on three key ingredients of the pressure-strain interaction: the normal and shear components of $Pi-D$ and pressure dilatation.
Our results demonstrate that the scale-filtered pressure–strain interaction is dominated by scale-filtered $Pi-D$ across the kinetic range, with the shear component consistently providing the dominant contribution. 
We find that the scale-filtered normal and shear contributions of $Pi-D$ exhibit persistent anticorrelation and opposite signs across all kinetic scales. 
We also discover that the amplitude of both anisotropic components for each species scales directly with their temperature and inversely with the temperature of the other species, while the scale-filtered pressure dilatation remains negligible compared to the $Pi-D$ terms but shows enhanced compressibility effects as plasma temperatures decrease.
We discuss the implications of these findings in thermally non-equilibrated plasmas, such as in the turbulent magnetosheath and solar wind.
\end{abstract}

\maketitle

\section{Introduction}
\label{sec:intro}
Plasmas are often weakly collisional in diverse naturally occurring environments, such as in the planetary magnetospheres, solar wind, and in other astrophysical systems. Because collisions are rare, these plasmas often depart significantly from local thermodynamic equilibrium (LTE), which obscures the mechanisms through which dissipation, \textit{i.e.}, irreversible conversion of energy, takes place during multiscale interconnected phenomena such as magnetic reconnection, turbulence, and collisionless shocks~\cite{Howes17,matthaeus2020pathways}.

Over the past decade, pressure–strain interaction has received considerable attention in the context of collisionless kinetic plasmas~\citep{yang2017energya, chasapis2018situ,bandyopadhyay2020statistics,pezzi2021dissipation,cassak2022pressure,burch2023electron}. Originally thought of as a channel for energy conversion between bulk flow and internal (thermal) energies~\cite{yang2017energyb}, it has since been increasingly evolved as a proxy for investigating ``dissipation-like" processes in systems lacking explicit closure models. In contrast to magnetohydrodynamics (MHD), where closure is achieved by modeling dissipation through viscosity and resistivity, non-LTE kinetic plasmas do not possess analogous fluid closures, since in weakly collisional plasmas, higher-order moments of the non-Maxwellian phase space density can be non-zero and important~\cite{Cassak_FirstLaw_2023,Barbhuiya_PRE_2024}. The pressure–strain interaction has emerged as a leading candidate to address this gap, supported by multiple theoretical frameworks. One such approach employs conditional averaging~\cite{bandyopadhyay2023collisional,yang2024effective}, wherein part of the pressure–strain interaction, when conditioned on a threshold of the traceless velocity strain rate tensor $\vct{D}$, scales as the trace of the squared velocity strain rate tensor $\vct{D}^2$. 
This result is analogous to the scaling seen for viscous dissipation in collisional plasma.
Since the Vlasov-Maxwell equations are formally time-reversible, they cannot capture classical dissipation that is irreversible. Nonetheless, in collisionless plasma turbulence, nonlinear interactions between large-amplitude, low-frequency electromagnetic fluctuations and particles can lead to chaotic orbits, and thus contribute to stochastic heating, which has been studied as a mechanism for irreversible dissipation\cite{Cerri_2021_ApJ}. Stochastic processes are found to compete with linear damping mechanisms, such as Landau damping, that lead to ``dissipation" through phase mixing of particle phase space densities\cite{Mallet_2019_JPP,Meyrand_2019_PNAS}. 
In the present study, we use the coarse-graining technique on Vlasov-Maxwell equations, which may introduce irreversibility by the nature of the technique itself which discards small-scale information, thus providing a different approach for understanding how irreversible dissipation arise in collisionless turbulence.

Using the coarse-graining technique\cite{germano1992turbulence}, it has been shown that in a Vlasov–Maxwell system, the total (ion plus electron) pressure–strain interaction accounts for the energy transfer at sub-ion (kinetic) scales for collisionless plasma processes such as turbulence~\cite{yang2022pressure} and reconnection~\cite{adhikari2024scale}. Quantification of energy transfer at these scales is notably absent in the von Kármán–Howarth (vKH)~\cite{de1938statistical,kolmogorov1991dissipation} formalism (based on structure functions), which is more commonly applied to study energy transfer in turbulent plasmas~\cite{politano1998dynamical, galtier2008karman, hellinger2018karman, ferrand2019exact}.

While vKH formalism is shown to be useful in estimating the energy cascade rate in kinetic collisionless plasma~\cite{hellinger2018karman, adhikari2021magnetic,yang2022pressure}, its validity is often limited by factors such as large-scale inhomogeneity, non-stationarity, lack of a well-defined inertial range, and finite Reynolds numbers. Especially in turbulent observational settings like the solar wind or the magnetosheath, general assumptions such as homogeneity and isotropy fail. In contrast, pressure-strain interaction as part of the scale filtering approach offers a more robust and physically complete measure of energy transfer at the kinetic scales, even in systems without isotropy and a clear inertial range. Furthermore, pressure-strain interaction for each species can be calculated, leading to the evaluation of each species' contribution to the thermal energy budget, which, therefore, is a useful tool to diagnose energy transfer in kinetic plasma.

Despite the progress made in understanding energy transfer at kinetic scales for a collisionless plasma, fundamental questions remain, such as the influence of the initial temperature of plasma species on the energy transfer at kinetic scales. This problem is at the core of the present study, where using kinetic particle-in-cell simulations of decaying plasma turbulence with the ratio of electron and ion temperatures being unequal to unity, we investigate how energy transfers at kinetic scales by applying the scale filtering technique to the pressure-strain interaction~\cite{yang2022pressure,manzini2022local,adhikari2024scale,hellinger2024anisotropy}.

This paper is organized as follows: Section~\ref{sec:theory} discusses the theoretical background, while in Section~\ref{sec:sim} we provide the details of the simulations. Section~\ref{sec:results} presents the results from the PIC simulations, and finally, in Section~\ref{sec:conclusions}, we conclude the study and discuss the implications.

\section{Review of the Theoretical Framework\label{sec:theory}}
Scale filtering~\citep{germano1992turbulence} or coarse-graining approach is a technique based on a well-defined filtering kernel $G_\ell = \ell^{-d}G(\vct{r}/\ell)$ such that the information of length scales $\geq \ell$ is preserved. Here, $G(\vct{r})$ is a normalized boxcar window function satisfying $\int d^d r G(\vct{r})=1$, with superscript $d$ being the number of dimensions of the system. For a field $f(\vct{x},t)$, the scale-filtered field quantity $\overline{f}_\ell(\vct{x},t)$ is defined as
\begin{equation}\label{eqn:scalefilter}
     \overline{f}_\ell(\vct{x},t) = \int d^d r G_\ell(\vct{r})f(\vct{x}+\vct{r},t).
\end{equation}

Likewise, we use the density-weighted filtered quantity $f(\vct{x},t)$, also called the Favre-filtered field~\citep{favre1969statistical,aluie2013scale} as
\begin{equation}\label{eqn:favrefilter}
     \tilde{f}_\ell(\vct{x},t) = \frac{\overline{\left[\rho(\vct{x},t) f(\vct{x},t)\right]}_\ell}{\overline{\rho}_\ell(\vct{x},t)},
\end{equation}
where $\rho(\vct{x},t)$ is the density. 

Using this scale filtering operation, we then spatially filter the Vlasov-Maxwell equations, compute the filtered moment equations, and obtain the time evolution of the filtered electromagnetic and the total bulk flow energy densities  (See ~\citet{matthaeus2020pathways} for details) as follows
\begin{eqnarray}
    \partial_t\overline{E}^m + \nabla \cdot \vct{J}^b &=& \Sigma_\alpha {\Lambda^{ub}_\alpha} - \Sigma_\alpha \Pi_\alpha^{bb} \label{eqn:em_field},\\
    \partial_t\Tilde{E}^f_\alpha + \nabla \cdot \vct{J}^u_\alpha &=& -\Pi_\alpha^{uu} - \Phi_\alpha^{uP} -\Lambda_\alpha^{ub}\label{eqn:u_field}, 
\end{eqnarray}
where $\alpha$ represents plasma species; $\overline{E}^m$ = $\frac{1}{8\pi} (\overline{\vct{B}}^2+\overline{\vct{E}}^2)$ is the filtered electromagnetic energy density with $\vct{B}$ and $\vct{E}$ being the magnetic and electric field respectively.
$\Lambda_\alpha^{ub}=-q_\alpha \overline{n}\Tilde{\vct{E}}\cdot \Tilde{\vct{u}}_\alpha$ represents the time rate of conversion of fluid flow energy density of species $\alpha$ into electromagnetic energy density, $\vct{u}_\alpha$ is filtered bulk flow velocity, and $\overline{n}_\alpha$ is the filtered number density.
$\vct{J}^b=c/4\pi (\overline{\vct{E}}\times \overline{\vct{B}})$ is the spatial transport term for the scale-filtered electromagnetic energy density;
$\Pi_\alpha^{bb}=-q_\alpha \overline{n}_\alpha \Tilde{\vct{\tau}}_\alpha^e \cdot \Tilde{\vct{u}}_\alpha$ represents the sub-grid scale flux term for electromagnetic energy across scales due to nonlinearities, where $\Tilde{\vct{\tau}}_\alpha^e = \Tilde{\vct{E}}-\overline{\vct{E}}$. Similarly, $\Tilde{E}_\alpha^f = \frac{1}{2}\overline{\rho}_\alpha \Tilde{\vct{u}}_\alpha^2$ is the filtered bulk flow energy density
$\vct{J}^u_\alpha= \Tilde{\vct{E}}_\alpha^f \Tilde{u}_\alpha + \overline{\rho}_\alpha \Tilde{\tau}_\alpha^u \cdot \Tilde{\vct{u}}_\alpha +\overline{\vct{P}}_\alpha \cdot \Tilde{\vct{u}}_\alpha$ is the spatial transport related to the bulk flow with $\vct{P}_\alpha$ being the pressure tensor; 
$\Pi_\alpha^{uu} = -(\overline{\rho}_\alpha \Tilde{\tau}^u_\alpha\cdot \nabla)\cdot \Tilde{\vct{u}}_\alpha - \frac{q_\alpha}{c}\overline{n}_\alpha\Tilde{\tau}_\alpha^b\cdot \Tilde{\vct{u}}_\alpha$ is the sub-grid scale flux term for bulk flow energy across scales due to nonlinearities, where $q_\alpha$ is the charge of plasma species $\alpha$, 
$\Tilde{\tau}^u_\alpha=\widetilde{\vct{u}_\alpha \vct{u}_\alpha}-\Tilde{\vct{u}}_\alpha \Tilde{\vct{u}}_\alpha$, $\Tilde{\tau}^b_\alpha=\widetilde{\vct{u}_\alpha\times \vct{B}}-\Tilde{\vct{u}}_\alpha\times \Tilde{\vct{B}}$. Note that $\widetilde{\vct{u}_\alpha\times \vct{B}}$ is the Favre-filtered form of $\vct{u}_\alpha\times \vct{B}$ and so on. Similarly,  $\Phi_\alpha^{uP}=-(\overline{\vct{P}}_\alpha\cdot \nabla)\cdot \Tilde{\vct{u}}_\alpha$ is the filtered pressure strain interaction that corresponds to the rate of conversion of flow into internal energy.

\begin{figure}
    \centering
    \includegraphics[width=1\linewidth]{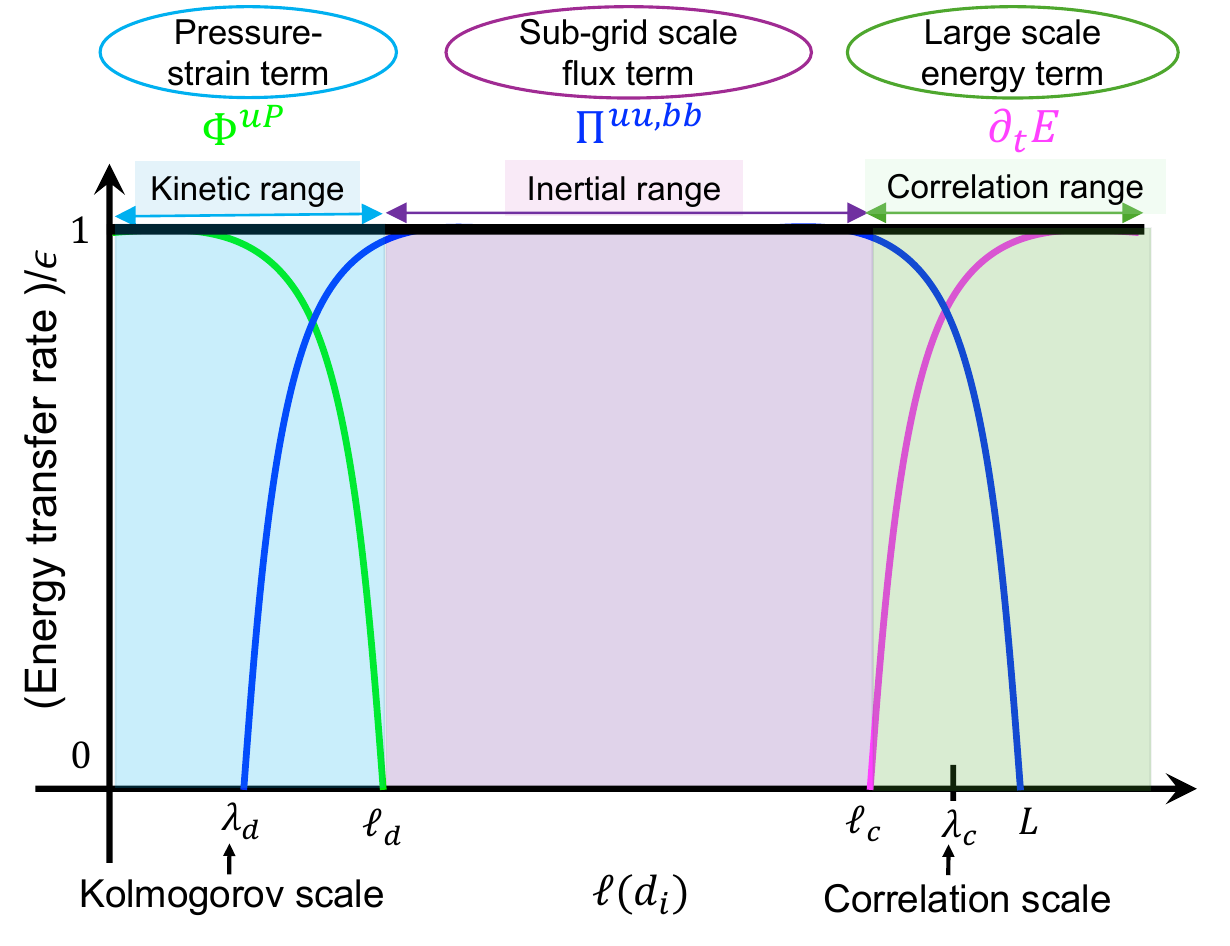}
    \caption{Idealized schematic plot of different terms of the scale filtered energy transfer equation (Eqn.~\ref{eqn:symbolic}) versus filtering width $\ell$. Each term (shown on top) is normalized to the total dissipation rate $\epsilon$, and different ranges where they dominate are shown by colored rectangles. Shaded rectangles of color blue, pink, and green represent the kinetic, inertial, and correlation range, respectively. Representative values of the correlation scale $\lambda_c$ and the Kolmogorov scale $\lambda_d$ are shown.}
    \label{fig:scalefiltered_plot}
\end{figure}

After an ensemble average, Eqs.~\ref{eqn:em_field} and~\ref{eqn:u_field}, can be combined to relate the energy transfer at the largest scale to the subgrid scale flux transport
\begin{equation}
\label{eqn:scalefiltered}
     \partial_t \bigl< \sum_\alpha \Tilde{E}^f_\alpha \! + \overline{E}^m \bigr> =  \! -\bigl< \sum_\alpha (\Pi^{uu}_\alpha \! + \Pi^{bb}_\alpha) \bigr> -\bigl< \sum_\alpha \Phi^{uP}_\alpha \bigr>,
 \end{equation}

In kinetic plasmas, where the exact form of dissipation may not be known, the total dissipation rate can be calculated by $\epsilon = -d\langle \sum_\alpha E_\alpha^f + E^m\rangle/dt=-\partial_t(\langle \sum_\alpha E_\alpha^f + E^m\rangle)-\vct{u} \cdot \nabla\langle \sum_\alpha E_\alpha^f + E^m\rangle$, where $\vct{u}$ is the bulk flow velocity of the plasma. Finally, we can rewrite Eqn.~\ref{eqn:scalefiltered} to provide a generalized picture of energy transfer as
\begin{equation}\label{eqn:symbolic}
     \partial_t E + \Pi^{uu,bb} + \Phi^{uP} = \epsilon.
\end{equation}

The first term on the left-hand side of Eq.~\ref{eqn:symbolic} is the local time rate of change of flow and magnetic energy for scales $\geq \ell$. This term vanishes for large enough $\ell$ and approaches $-\epsilon$ as $\ell \xrightarrow{} 0$. With this property, one can define the time rate of change of energy for scales $< \ell$ as $\partial_t E=\epsilon + \partial_t \langle \sum_\alpha \Tilde{E}^f_\alpha +\overline{E}^m \rangle$ such that $\partial_t E \xrightarrow{}\epsilon$ at large $\ell$ and $\partial_t E \xrightarrow{}0$ as $\ell \xrightarrow{}0.$ The second term defined as $\Pi^{uu,bb}=-\Sigma_\alpha (\Pi_\alpha^{uu}+\Pi_\alpha^{bb})$ is associated with the nonlinear energy flux, and the third term  $\Phi^{uP}_\alpha$, represents the internal energy deposition due to the pressure strain interaction, which is further decomposed as
\begin{equation}\label{eqn:ps_decomposition}
    \Phi_\alpha^{uP} = \sum_\alpha \langle -(\overline{\vct{P}}_\alpha\cdot \nabla) \cdot \Tilde{\vct{u}}_\alpha \rangle
    =\sum_\alpha \langle -\overline{p}_\alpha \nabla \cdot \Tilde{\vct{u}}_\alpha \rangle - \langle \overline{{\Pi}}_\alpha : \Tilde{\vct{D}}_\alpha \rangle,
\end{equation}
where $p_\alpha=P_{\alpha,ii}/3$ is the scalar effective pressure, $\Pi_{\alpha,ij}=P_{\alpha,ij}-p_{\alpha}\delta_{ij}$ is the \textit{ij} element of deviatoric pressure tensor, $\delta_{ij}$ is the Kronecker delta and $D_{\alpha,ij}=(\partial_i u_{\alpha,j} + \partial_j u_{\alpha,i})/2-\nabla \cdot \vct{u}_{\alpha} \delta_{ij}/3$ is the \textit{ij} element traceless strain rate tensor are the terms before scale filtering.

For brevity, we suppress the species subscript $\alpha$ from here on, only using it when necessary to avoid confusion.
With the recent deformation-based 
decomposition~\cite{cassak2022pressure} of 
$D$ into $D_{normal}$ and $D_{shear}$, we can split the last term of Eqn.~\ref{eqn:ps_decomposition} as 
\begin{equation}\label{eqn:ps_decomposition_final}
    \Phi^{uP} = \underbrace{\langle -\overline{p} \nabla \cdot \Tilde{\vct{u}} \rangle}_{\langle \overline{p}\tilde{\theta}\rangle} +\underbrace{ \langle -\overline{{\Pi}} : \Tilde{\vct{D}}_{shear}\rangle}_{\langle \overline{Pi}-\tilde{D}_{shear}\rangle} +\underbrace{ \langle -\overline{{\Pi}} : \Tilde{\vct{D}}_{normal} \rangle}_{\langle \overline{Pi}-\tilde{D}_{normal}\rangle}.
\end{equation}

Unlike other equations based on MHD models, Eqn.~\ref{eqn:symbolic} is entirely derived from the Vlasov-Maxwell model and therefore is an ideal candidate to describe its energy characteristics. Under certain assumptions, Eqn.~\ref{eqn:symbolic} has been shown to be analogous to the von K\'arm\'an-Howarth equation~\citep{de1938statistical,monin1975statistical},  derived through a completely different formalism based on structure functions. Note that both these formalisms describe the conservation of energy across different scales and are composed of different energy transfer terms, while there exists a correspondence between individual terms of Eqn.~\ref{eqn:symbolic} and the terms in the vKH 
equation.
$\partial_t E$ is equivalent to the time rate of change of energy within a lag $\ell$, $\Pi^{uu,bb}$ is equivalent to the nonlinear energy transfer dominant in the inertial range, and $\Phi^{uP}$ is equivalent to the visco-resistive dissipation in the MHD description. In a fully developed turbulence cascade, different filtered terms of Eqn.~\ref{eqn:symbolic} are expected to dominate at different length scales: the time derivative term $\partial_t E$ reaches the total dissipation rate $\epsilon$ at scales larger than the correlation length, decreases at intermediate scales (roughly the inertial range) where $\Pi^{uu,bb}$ dominates. 
At the smallest scales ($\ell \xrightarrow{} 0$), the rate of production of internal energy $\Phi^{uP}$
becomes dominant and accounts for the total dissipation. This technique has been extensively applied to kinetic simulations of turbulence~\cite{yang2022pressure} and reconnection~\cite{adhikari2024scale} and Magnetospheric Multiscale (MMS)  observations\cite{roy2025scalefilter}.

Fig.~\ref{fig:scalefiltered_plot} shows an idealization of the variation of different terms in Eqn.~\ref{eqn:symbolic} as a function of the filtering scale $\ell$(lag space). A reference correlation scale $\lambda_c$ and Kolmogorov scale $\lambda_d$ are set for convenience. Note that each term is normalized by the mean rate of change of flow and electromagnetic energy, and the sum at any lag would correspond to the total dissipation rate. At the larger lag scales (energy containing or correlation range), the time rate of change of scale filtered energy $\partial_t E$ in the entire system equals the total dissipation, $\partial_t E = -\epsilon$. At the intermediate scales, the influences of $\partial_t E$ and $\Phi^{uP}$ terms are small, and the inertial range is dominated by the sub-grid scale flux terms $\Pi^{uu,bb}$. At the smaller kinetic scales ($\ell < \ell_d$), the filtered pressure-strain interaction $\Phi^{uP}$ dominates over the other terms. A similar term-by-term analysis for the vKH equation is presented in~\citet{adhikari2023effect}.

At this point, we remind the reader that the analysis based on the scale-filtered energy equation is analogous to the structure function-based vKH approach. This correspondence has been demonstrated in detail for simulations of kinetic turbulence~\cite{yang2022pressure}. While the overall behavior of the analogous terms in both approaches is consistent across their respective length scales (kinetic, inertial, and correlation ranges), some discrepancies arise in the delineation of these ranges. For instance, the scale filtering method tends to overestimate the extent of the kinetic range compared to the structure function-based vKH approach, resulting in a slight shift of the inertial and energy-containing ranges by a few ion inertial lengths ($d_i$). However, for this study, this discrepancy is not a concern since we are focusing on the kinetic range $\ell <d_i$, where both the approaches show strong agreement.

In this paper, we focus on the smallest lag-scale, or in other words, the kinetic range, and investigate how the ion and electron thermal disequilibrium, \textit{i.e.,} unequal initial scalar effective temperatures of the species ($T_i\neq T_e$), where $T_\alpha =p_\alpha/(n_\alpha k_B)$ and $k_B$ is the Boltzmann constant, influences the energy budget.

\section{Simulations \label{sec:sim}}
We employ a massively parallel particle-in-cell code {\tt p3d} \cite{zeiler2002three} to perform numerical simulations of decaying turbulence. All simulations performed are 3D in velocity-space, and 2.5D in position-space, meaning that vectors have three components and there is one invariant spatial dimension, which for these simulations is the out-of-plane ($\hat{z}$) direction.
{\tt p3d} uses a relativistic Boris particle stepper \cite{Birdsall85} to evolve macro-particles and the trapezoidal leapfrog method \cite{guzdar93a} to evolve electromagnetic fields forward in time.
{\tt p3d} uses the multigrid method \cite{Trottenberg00} to clean the electric field and enforce Poisson's equation.
For all simulations, we employ periodic boundary conditions in both spatial directions.

All simulated quantities are output in code-normalized units. 
Magnetic fields are normalized to a reference magnetic field $B_0$. 
Times are normalized to the inverse ion cyclotron frequency $\omega_{ci}^{-1}= (q_i B_{0} / m_{i} c)^{-1}$, where $c$ is the speed of light, and $q_i$ and $m_i$ are the ion charge and mass, respectively.
Lengths are normalized to the ion inertial scale $d_{i} = c/\omega_{pi0}$, where $\omega_{pi0} = (4 \pi n_0 q_i^2 /m_i)^{1/2}$ is the ion plasma frequency based on a reference number density $n_0$. 
Velocities are normalized to the ion Alfv\'en speed $c_{A0} = B_0/(4 \pi m_i n_0)^{1/2}$. 
Temperatures are normalized to $m_i c_{A0}^2/k_B$.
Lastly, power densities are presented in units of $(B_0^2/4 \pi) \Omega_{ci0}$.

We employ unrealistic values of the speed of light $c=15$ and electron-to-ion mass ratio $m_e/m_i = 0.04$ for numerical expedience, but we expect these values do not qualitatively change the results presented here. 
A square and periodic domain of size $L_x \times L_y = 37.3912 \times 37.3912$ with 1024 x 1024 grid cells initialized with 6,400 particle-per-grid (PPG) for all five simulations. The choice of high PPG is advantageous in reducing PIC noise at small length scales being studied in the present work. We use an initial uniform magnetic field that has a strength given by the reference value $B_0=1$ and points along $\hat{z}$. We seed the initial system with Alfv\`enic velocity and magnetic field fluctuations with random phase that are excited in a band of wave numbers $k \in [2,4] \times 2 \pi/L_x$ with a flat spectrum.  The initial root mean square fluctuation amplitude for both velocity and magnetic field is set to 0.25, and thus, the initial cross helicity is very small (less than 0.1). The initialization conditions of the five simulations follow those used in a previous study, where the domain size is four times larger ~\cite{parashar2018dependence}.
For our simulations, the time scale over which non-linear interactions between different wave modes become significant for the system-size length-scale structures, called the non-linear time, is $\tau_{nl} = L_x/[2 \pi (\delta b_{rms}^2 + \delta v_{rms}^2)^{1/2}] \simeq 17$, and we evolve the systems for approximately 4$\tau_{nl}$.

At initialization, a uniform number density is used that is set by the reference density $n_0=1$.
Each species, \textit{i.e.,} electrons and ions, is initialized with drifting Maxwellian velocity distribution functions at the initial uniform temperatures, $T_{e}$ and $T_{i}$, respectively, and they drift at the local bulk flow speed. 
The smallest time scale of the system in the five simulations is the inverse of electron plasma frequency $\omega^{-1}_{pe} \simeq 0.0133$, and we choose the particle time-step to be $\Delta t=0.005$ with the field time step $\Delta t/3$.
The five simulations have varying $T_{e}/T_{i}$ ratio and the key differences in the simulations are listed in Table \ref{tab:simparameters}, where the third column lists the smallest characteristic kinetic length scale and whether it is the electron Debye ($\lambda_{De}$) or ion Debye ($\lambda_{Di}$) scale for the particular $T_{e}/T_{i}$ simulation.
For the five simulations, the grid length in both spatial directions is set to $\Delta=0.03651$, and as noted in the Table \ref{tab:simparameters}, it means that for the simulations with $T_{e}/T_{i}=0.25$ and 4, the smallest length scale is not resolved, for $T_{e}/T_{i}=0.5$ and 2,  the smallest length scale is the grid length,  and $T_{e}/T_{i}=1$, the smallest length scale is well-resolved. 
The electric field is cleaned every 40 particle time steps to ensure Poisson's equation is satisfied and improve energy conservation, 
which we find is within 0.1\% for the five $T_e/T_i$ simulations by the final time of $t=75$.

\begin{table}
    \centering
    \begin{tabular}{|c|c|c|c|}
    \hline
       $T_e$  & $T_i$ & $T_e/T_i$ & smallest length scale \\
    \hline 
        0.15 & 0.60 & 0.25 & 0.02582 ($\lambda_{De}$)\\
        0.30 & 0.60 & 0.5 & 0.03651 ($\lambda_{De}$)\\
        0.60 & 0.60 & 1.0 & 0.05164 ($\lambda_{De}~\text{\&}~\lambda_{Di}$)\\
        0.60 & 0.30 & 2.0 & 0.03651 ($\lambda_{Di}$)\\
        0.60 & 0.15 & 4.0 & 0.02582 ($\lambda_{Di}$)\\
        \hline
    \end{tabular}
    \caption{Electron ($T_e$) and ion ($T_i$) temperatures at initialization are listed for the five turbulence simulations, with their temperature ratio ($T_e/T_i$) in the third column. The fourth column presents the smallest length scale and whether it is the electron or ion Debye length.}
    \label{tab:simparameters}
\end{table}

Before analyzing the scale-filtered pressure–strain interaction, we mitigate the effects of PIC noise by applying a recursive smoothing procedure to the raw simulation data. Specifically, the pressure tensor and velocity fields for each species are smoothed over a two-cell width, repeated three times. This smoothing width is chosen to preserve information at the smallest filtering width, and its appropriateness is verified by comparing the scale-filtered pressure–strain interaction at that scale.
We focus on studying the time evolution of the decomposition of the pressure-strain interaction for both ions and electrons over a time interval of $[7,51]\omega_{ci}^{-1}$. During the initial phase ($t\omega_{ci}<7$), these systems undergo transient fluctuations associated with the initial conditions. As a result, almost all the pressure-strain elements have oscillatory behavior. At late stages, $t\omega_{ci}>51$, for a decaying turbulence, the mean square current starts to fall. Therefore, for the analysis, we only use the intermediate times.

\section{Results\label{sec:results}}
We first qualitatively discuss the evolution of the decomposition of the scale filtered pressure-strain interaction (see Eq.~\ref{eqn:ps_decomposition_final}) as a function of lag for all the simulations.

\begin{figure*}
    \centering
    \includegraphics[width=1.\linewidth]{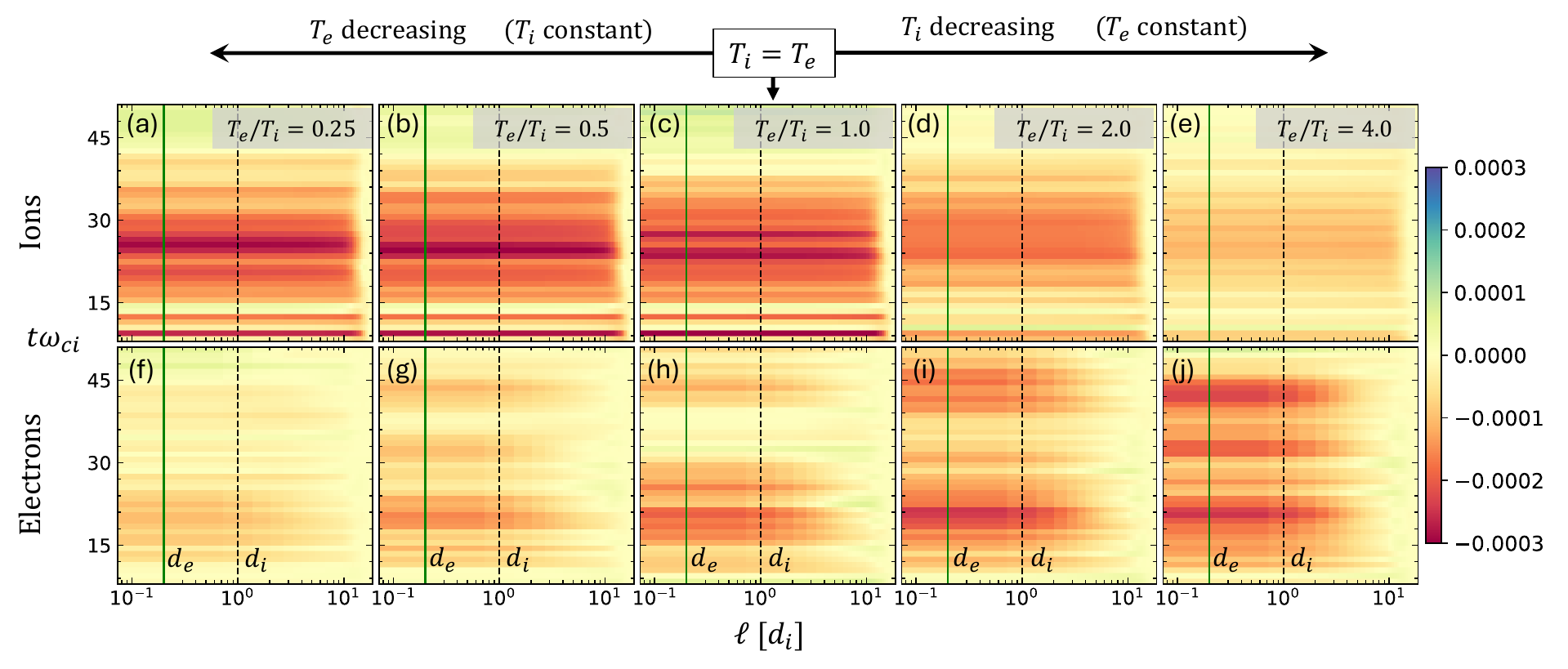}
    \caption{A comparison of time evolution of scale-filtered domain averaged $\langle \overline{Pi}-\tilde{D}_{normal} \rangle$ for ions in panels (a)-(e) and electrons in panels (f)-(j) as a function of lag $\ell$ in systems with different electron to ion temperature ratios. A dashed line at $d_i$ and a green solid line at $d_e$ are drawn for reference.}
    \label{fig:pidnormal_final}
\end{figure*}

\begin{figure*}
    \centering
    \includegraphics[width=1.\linewidth]{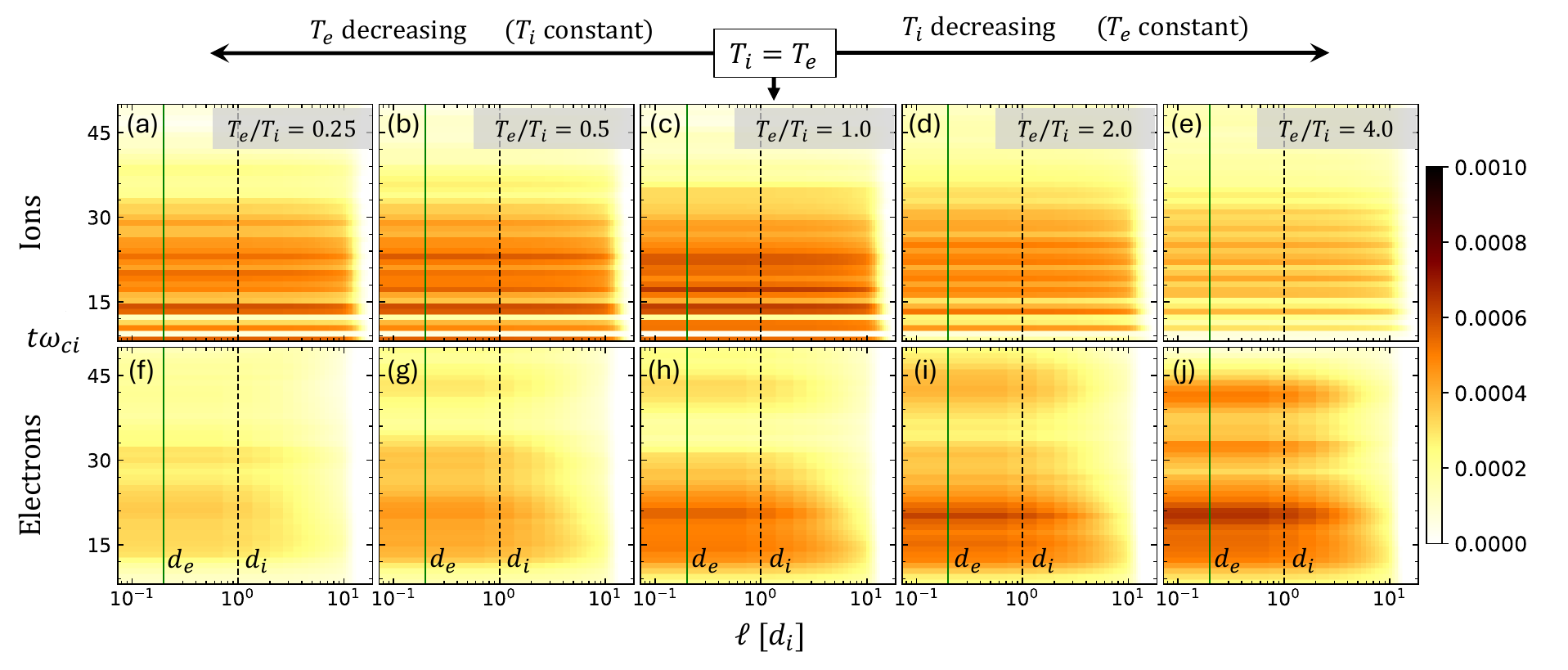}
    \caption{A comparison of time evolution of scale-filtered domain averaged $\langle Pi-D_{shear} \rangle$ for ions in panels (a)-(e) and electrons in panels (f)-(j) as a function of lag $\ell$. A dashed line at $d_i$ and a green solid line at $d_e$ are drawn for reference. Note that the colorbar is different than the one in Fig.~\ref{fig:pidnormal_final} (See text for details).
    \label{fig:pidshear_final}
    }
\end{figure*}

\begin{figure*}
    \centering
    \includegraphics[width=1.\linewidth]{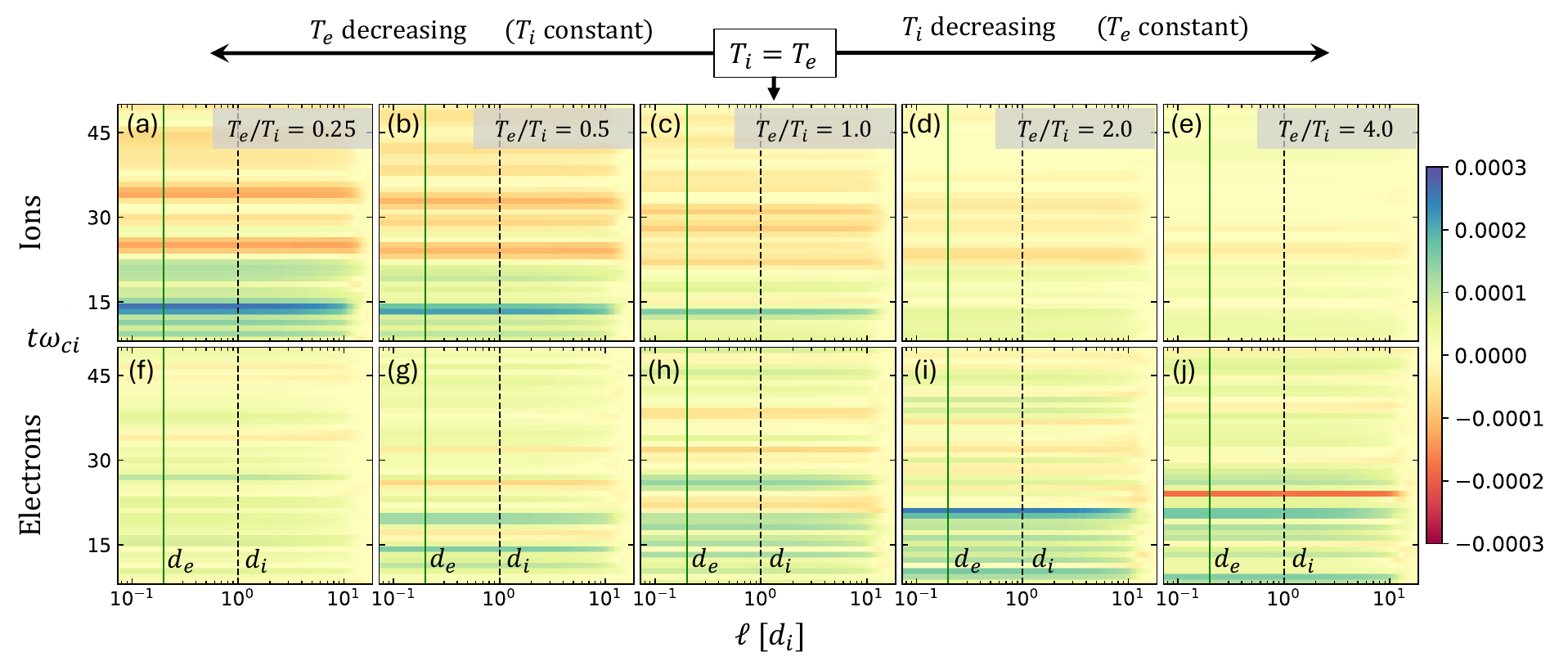}
    \caption{A comparison of time evolution of scale-filtered domain averaged $\langle \overline{p}\tilde{\theta} \rangle$  for ions in panels (a)-(e) and electrons in panels (f)-(j) as a function of lag $\ell$. A dashed line at $d_i$ and a green solid line at $d_e$ are drawn for reference.}
    \label{fig:ptheta_final}
\end{figure*}

In Fig.~\ref{fig:pidnormal_final}, \ref{fig:pidshear_final} and \ref{fig:ptheta_final}, the first two columns represent cases where  $T_i$ at initialization is kept constant and $T_e$ increases from left to right, while the last two columns represent the cases where $T_e$ at initialization is kept constant and $T_i$ decreases from left to right, as shown by the arrows on the top of all three figures. 
The central column in each figure represents the case with $T_e=T_i$  at initialization. 
In each figure, we demonstrates data for ions (top row, panels (a)-(e)) and electrons (bottom row, panels (f)-(j)) as a function of lag, across a range of electron to ion temperature ratios which vary from $0.25$ to $4$ as we move from first column to the fifth. 
A green (solid) vertical line represents the electron inertial length $d_e$ while the black (dashed) line represents the ion inertial length $d_i$.

We start our discussion with the time evolution of 
$\langle \overline{Pi}-\Tilde{D}_{normal}\rangle$ as shown in Fig.~\ref{fig:pidnormal_final}.
Looking at the ions, we find that 
when $T_i \geq T_e$  (i.e., $T_e/T_i \leq 1$), oscillatory patches of 
$\langle \overline{Pi}-\Tilde{D}_{normal}\rangle$ are present after initialization of the simulations.  
As the system evolves, the mean square current density increases (not shown), and as it peaks (around $\sim 21$), turbulence is fully developed. During this phase ($t\omega_{ci}\sim 20$), significant amplitudes of $\langle \overline{Pi}-\Tilde{D}_{normal}\rangle$
appear around the electron and ion scales ($\ell \sim d_i$) which last for about $tw_{ci}=15-20$.
These structures weaken as the electron temperature increases, seen when moving to the left from (a) to (c) or ion temperature decreases, seen as we move to the left from (c) to (e), indicating reduced ion-scale activity in high $T_e$ cases. Conversely, for electrons 
$\langle \overline{Pi}-\Tilde{D}_{normal}\rangle$ is significant mostly at the electron scales ($\ell\sim d_e$) and persists to later times ($t\omega_{ci}\sim 30 - 45$) as the ratio $T_e/T_i$ increases (when moving to the left from (f) to (j)). The strongest electron-scale activity is observed for the case with the strongest initial disequilibration, \textit{i.e.}, $T_e/T_i=4$ (panel (j)), suggesting a transition in the dominance of electron
$\langle \overline{Pi}-\Tilde{D}_{normal}\rangle$ over ions with increasing $T_e$. While we do observe positive values of $\langle \overline{Pi}-\Tilde{D}_{normal}\rangle$ 
either at larger lags or at late times, it is consistently negative at the sub-ion scales $\ell < d_i$ for both ions and electrons.

In Fig.~\ref{fig:pidshear_final}, we show the time evolution of
$\langle \overline{Pi}-\Tilde{D}_{shear}\rangle$,
once again as a function of lag, across these simulations.
For ions (panels (a)-(e)), $\langle \overline{Pi}-\Tilde{D}_{shear}\rangle$ exhibits persistent oscillatory bands of relatively significant amplitudes. Similar oscillatory trends have been observed in decaying~\cite{hellinger2024anisotropy},  and shear~\cite{goodwill2025nonlinear} turbulence, and can be attributed to the initialization based on magnetohydrodynamic (MHD) conditions~\cite{cerri2013extended}. Also note that we have now used a different colorbar with no negative values. This is because $\langle \overline{Pi}-\Tilde{D}_{shear}\rangle$ is found to be positive in the kinetic range (which is the focus of this study) and sometimes negative only at the larger lags. As seen in Fig.~\ref{fig:pidshear_final} for constant $T_i$, ion 
$\langle \overline{Pi}-\Tilde{D}_{shear}\rangle$ increases slightly, extending towards larger scales $\ell>d_i$, with an increase in $T_e$ (as seen when moving leftward from panel (a) to (c)). This trend can be attributed to 
the increase in the total internal energy of the plasma 
when $T_e$ increases, while $T_i$ is kept fixed.
Furthermore, as $T_i$ is decreased while 
keeping $T_e$ constant (seen when we move leftward from panel (c) to (e)), small-scale $\langle \overline{Pi}-\Tilde{D}_{shear}\rangle$ 
diminishes, suggesting reduced ion-scale features.
In contrast, electron (panels (f)-(j)) shows a markedly different behavior. At low $T_e/T_i$ (panels (f),(g)), when ions are hotter than the electrons at initialization, electron $\langle \overline{Pi}-\Tilde{D}_{shear}\rangle$  
is weaker in amplitude and more confined in both lag scale to ion-scales and below, and in time. As $T_e/T_i$ increases (when moving leftward from panel (f) to (j)), $\langle \overline{Pi}-\Tilde{D}_{shear}\rangle$  
intensifies and broadens in both lag space and in time, particularly for scales near and just below $d_i$. This broadening is likely  
related to the enhanced energy transfer at electron scales, as electron internal (thermal) energy 
increase. In the two cases with $T_e > T_i$ (shown in panels (i),~(j)), we find that there exist bands of elevated electron $\langle \overline{Pi}-\Tilde{D}_{shear}\rangle$ 
close to the electron inertial scales $\ell\sim 0.3 d_i$, possibly related to electron-scale fluctuations and transition to electron dissipation mechanisms found through the scale-filtered formalism~\cite{adhikari2025estimation}. It is worth mentioning that $\langle \overline{Pi}-\Tilde{D}_{shear}\rangle$ 
at kinetic scales is always positive while $\langle \overline{Pi}-\Tilde{D}_{normal}\rangle$ 
is always negative. This feature 
is also observed in $\langle \overline{Pi}-\Tilde{D}_{normal}\rangle$ and $\langle \overline{Pi}-\Tilde{D}_{shear}\rangle$  when analyzed as a cumulative running average or as a domain average as a function of time 
in simulations of turbulence with varying $T_e/T_i$ \cite{Barbhuiya_2026_PoP}
and reconnection~\cite{adhikari2025_PSguidefield}.

Having discussed the anisotropic contribution to the scale-filtered pressure-strain interaction, we move on to examine the remaining isotropic compressional part, so-called pressure dilatation $\bar{p}\Tilde{\theta}$.
In Fig.~\ref{fig:ptheta_final}, we compare and contrast the time evolution of $\langle \bar{p}\Tilde{\theta}\rangle$ 
as a function of lag. For ions, at low $T_e$ while $T_i$ is fixed (shown in panels (a), (b)),
the ion $\langle \bar{p}\Tilde{\theta}\rangle$ 
exhibits pronounced positive values at $t\omega_{ci}\sim 15$ for all lag scales 
due to compression in the initial phase. 
For $t\omega_{ci} \lesssim 36$, ion $\langle \bar{p}\Tilde{\theta}\rangle$ exhibit negative values, 
due to expansion at those lag scales. As $T_e$ approaches $T_i$ (panel (c)), these features start to weaken and become short-lived. 
For
high $T_e/T_i$ cases (panels (d) and (e)), the ion 
$\langle \bar{p}\Tilde{\theta}\rangle$ becomes more diffuse and largely less negative 
due to the enhancement of 
the electron dynamics, which in turn compresses colder ions.
Electrons, on the other hand (panels (f)-(j)) exhibit increasingly structured behavior in $\langle \bar{p}\Tilde{\theta}\rangle$ with increasing $T_e/T_i$, particularly at electron kinetic scales ($\ell \lesssim d_e$) in the initial phase of turbulence. 
For low $T_e/T_i$ cases (panels (f), (g)), electron 
$\langle \bar{p}\Tilde{\theta}\rangle$ is minimal and shows only faint scale-dependent features, indicating weak compression of the electron. As $T_e/T_i$ increases (when we move leftward from (h)), distinct 
oscillating bands of $\langle \bar{p}\Tilde{\theta}\rangle$ emerge near $\ell \sim 0.5\text{–}1\,d_e$,
suggestive of enhanced electron compression and expansion effects at small lag scales. 

\begin{figure*}
    \centering
    \includegraphics[width=1\linewidth]{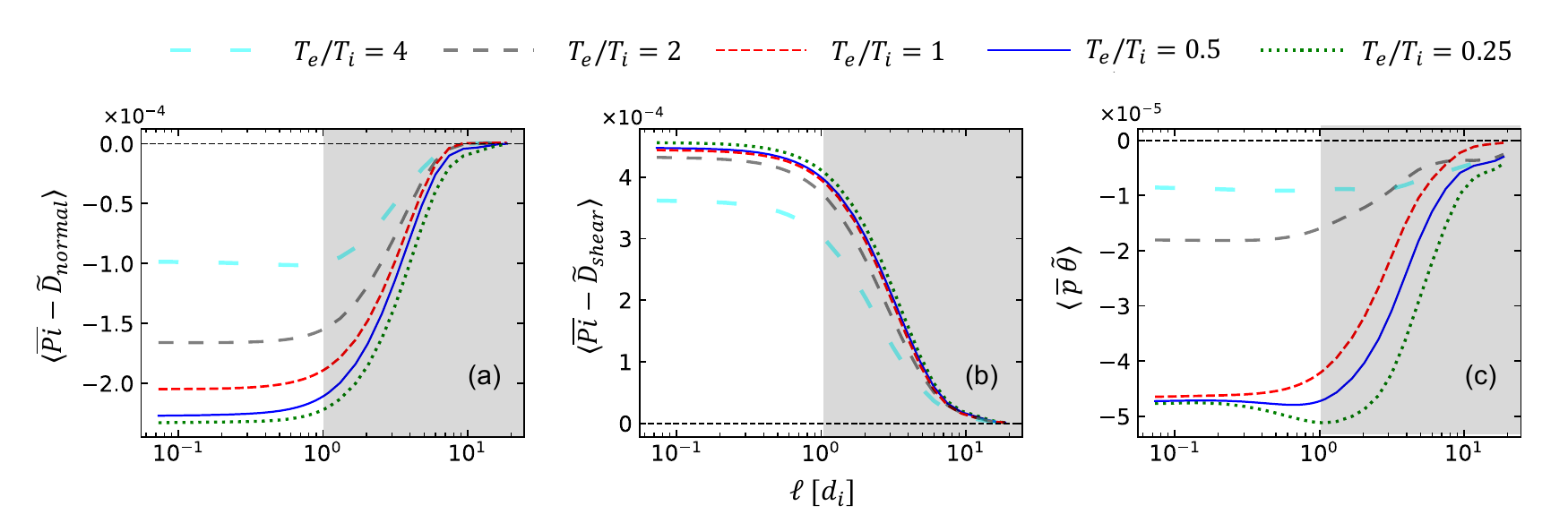}
    \caption{Comparison of the scale dependence of ingredients of ion pressure-strain interaction: (a) $\langle \overline{Pi}-\tilde{D}_{normal} \rangle$, (b) $\langle \overline{Pi}-\tilde{D}_{shear} \rangle$ and (c) pressure dilatation $\langle \overline{p} \tilde{\theta} \rangle$ for cases with different electron to ion temperature ratio.}
    \label{fig:ps_ions}
\end{figure*}

\begin{figure*}
    \centering
    \includegraphics[width=1\linewidth]{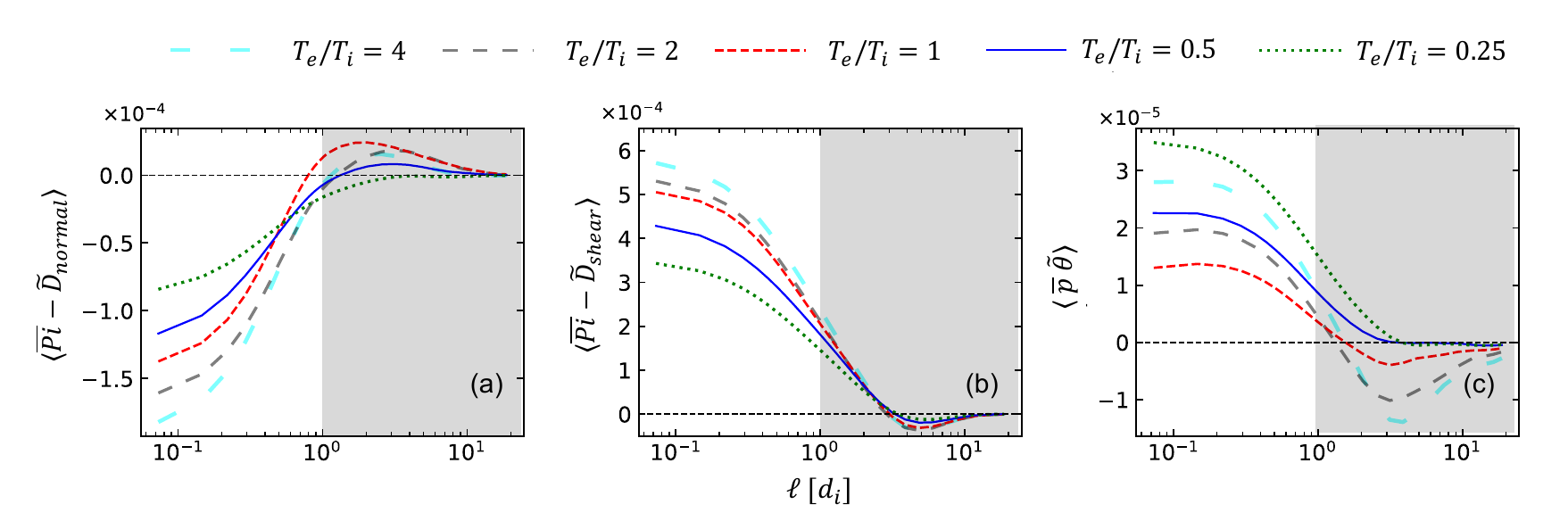}
    \caption{Comparison of the scale dependence of ingredients of electron pressure-strain interaction: (a) $\langle \overline{Pi}-\tilde{D}_{normal} \rangle$, (b) $\langle \overline{Pi}-\tilde{D}_{shear} \rangle$ and (c) pressure dilatation $\langle \overline{p} \tilde{\theta} \rangle$ for cases with different electron to ion temperature ratio.}
    \label{fig:ps_electrons}
\end{figure*}

Now that we have explored the time evolution of the scale-filtered domain-averaged pressure-strain interaction ingredients and found that the energy transfer at the kinetic scale is dominated by the (scale-filtered) pressure-strain interaction of that species (ion or electron) whose temperature is much larger than the other species. In fact, the contribution of pressure-strain interaction is dominated by $\langle \overline{Pi}-\tilde{D}_{shear} \rangle$ across the whole kinetic scale. Next, we detail the differences between these pressure-strain elements in the kinetic range. For that, our focus is on times immediately after the mean square current density peaks and contrasting the differences with respect to lag. 
We note that while the pressure-strain interaction dominates at kinetic scales, it becomes suppressed relative to the flux term in the inertial range and the energy derivative terms at the largest (energy-containing) scales. Consequently, our analysis focuses on the lag scales smaller than the ion inertial length $\ell \sim d_i$, where these effects are most pronounced. The shaded rectangle in each panel of Fig.~\ref{fig:ps_ions} and \ref{fig:ps_electrons} is therefore outside of the region of interest. To do so, we average the data shown in Fig.~\ref{fig:pidnormal_final}-\ref{fig:ptheta_final} for $5$ time slices, from
$t\omega_{ci}=25$ through $t\omega_{ci}=30$ and compare the various pieces that make up pressure-strain interaction.
Fig.~\ref{fig:ps_ions} shows the three pieces of the ion pressure-strain interaction for different electron to ion temperature ratio $(T_e/T_i)$. A wide dashed cyan line represents $T_e/T_i = 4$, a dashed gray line denotes $T_e/T_i = 2$, a small dashed red line corresponds to $T_e/T_i = 1$, a solid blue line indicates $T_e/T_i = 0.5$, and a dotted green line represents $T_e/T_i = 0.25$ case.
With $T_e$ fixed, as the ion temperature decreases alongside the increase in the ratio of electron to ion temperature (cases with $T_e/T_i=1, 2,$ and $4$)
both $\langle \overline{Pi}-\Tilde{D}_{normal}\rangle$
(panel (a)) and 
$\langle \overline{Pi}-\Tilde{D}_{shear}\rangle$  (panel (b)) exhibits a reduction in amplitude across sub-ion scales. This reduction is more pronounced in  $\langle \overline{Pi}-\Tilde{D}_{normal}\rangle$,
which decreases by approximately a factor of $2$ when $T_e/T_i$ increases from $1$ to $4$. In contrast, the decline in  $\langle \overline{Pi}-\Tilde{D}_{shear}\rangle$ is less drastic over the same range. For cases where ion temperature is constant and electron temperature decreases ($T_e/T_i=0.5,$ and $0.25$), both quantities show no clear trend for $\ell \lesssim 2d_e$, and 
$\langle \bar{p}\Tilde{\theta}\rangle$ stays roughly constant at scales $\ell < 2d_e$ but increases in amplitude for larger lags $\ell > 2d_e$. Moreover, ion $\langle \bar{p}\Tilde{\theta}\rangle$ is observed to become less negative as $T_i$ decreases, indicating enhanced compressibility in colder ions (panel (c)).

Lastly, in Fig.~\ref{fig:ps_electrons}, we compare the electron pressure-strain interaction ingredients for all the cases of temperature ratios.
For the cases where $T_e$ is kept fixed, while $T_e \geq T_i$, we observe that $\langle \overline{Pi}-\Tilde{D}_{normal}\rangle$ becomes more negative as $T_i$ is decreased.
When $T_e$ decreases while keeping $T_i$ fixed, the amplitude of electron $\langle \overline{Pi}-\Tilde{D}_{normal}\rangle$ also decreases, becoming less negative as shown in Fig.~\ref{fig:ps_electrons}(a).
Interestingly, a crossover in behavior emerges near $\ell \sim 0.6d_i$, after which the electron $\langle \overline{Pi}-\Tilde{D}_{normal}\rangle$ 
becomes positive for all cases except $T_e/T_i=0.25$, which remains negative across all lag values greater than $0.6d_i$ but approaches zero at larger scales.
Looking at Fig.~\ref{fig:ps_electrons}(b), we discover that when $T_e$ is fixed and $T_i$ is decreased, \textit{i.e.,} cases with $T_e>T_i$, the amplitude of electron $\langle \overline{Pi}-\Tilde{D}_{shear}\rangle$ increases. However, as $T_e$ decreases (with $T_i$ constant), the amplitude of electron $\langle \overline{Pi}-\Tilde{D}_{shear}\rangle$ diminishes, reflecting an overall weakening of deformation components of pressure-strain interactions in colder electron species.
Contrasting this trend, the electron $\langle \bar{p}\Tilde{\theta}\rangle$ (panel (c)) intensifies with a decrease in electron temperature. This behavior mirrors the ion response observed earlier, wherein cooler plasma conditions lead to increased compressibility~\cite{adhikari2025revisiting} and, consequently, a stronger $\langle \bar{p}\Tilde{\theta}\rangle$ signature.
At larger lags, however, the compressional effects on electrons are reduced.
For cases where $T_e$ is fixed, and $T_i$ is decreased, we find that electron $\langle \bar{p}\Tilde{\theta}\rangle$ increases. This is most likely because a decrease in $T_i$ leads to increased compressional effects for ions (see Fig.~\ref{fig:ps_ions}), which in turn will also compress electrons because of their relatively small mass.

\section{Conclusions and Future Work \label{sec:conclusions}}
In this paper, we use kinetic particle-in-cell simulations of decaying turbulence with varying electron-to-ion temperature ratios, where $\beta \sim 1$, to investigate the energy budget beyond the sub-ion scale, with a focus on the kinetic range. We present a generalized picture of energy transfer across scales using the scale filtering approach and explore the effect of changes in ion and electron temperature on the 
scale filtered pressure strain interaction.

The key findings of this paper are summarized  
below:

\begin{enumerate}
    \item We find that for higher thermal disequilibrium ($T_e \gg T_i$ or $T_e \ll T_i$ ), the scale-filtered pressure-strain interaction of the hotter plasma species dominates over the colder one. Further, we show that the dominant contribution to the scale-filtered pressure-strain interaction at the sub-ion scales comes from $\langle \overline{Pi}-\Tilde{D}_{shear}\rangle$. This highlights the importance of flow shear in describing the energy transfer at the kinetic scales.

    \item For individual species, the net amplitude of the anisotropic ingredients of the scale-filtered pressure-strain interaction is directly proportional to their temperatures. In particular, for a two species plasma, we show that the amplitude of both $\langle \overline{Pi}-\Tilde{D}_{normal}\rangle$ and $\langle \overline{Pi}-\Tilde{D}_{shear}\rangle$ for a species increases when the temperature of the other species decreases. For instance, ion $\langle \overline{Pi}-\tilde{D}_{normal} \rangle$ and $\langle \overline{Pi}-\tilde{D}_{shear} \rangle$ increases with 
    increase in $T_i/T_e$), while the electron $\langle \overline{Pi}-\tilde{D}_{normal} \rangle$ and $\langle \overline{Pi}-\tilde{D}_{shear} \rangle$ increases with increase in $T_e/T_i$.

    \item We find that across all kinetic range, $\langle \overline{Pi}-\Tilde{D}_{normal}\rangle$ and $\langle \overline{Pi}-\Tilde{D}_{shear}\rangle$ exhibit opposite signs for both species. While $\langle \overline{Pi}-\Tilde{D}_{shear}\rangle$ is positive, $\langle \overline{Pi}-\Tilde{D}_{normal}\rangle$ is negative for both ions and electrons.

    \item In the kinetic range, we find that the amplitude of the isotropic ingredient of the scale filtered pressure-strain interaction $\langle \overline{p} \tilde{\theta} \rangle$ increases with decreases in temperature for both species. This implies enhanced compressibility effects with a decrease in plasma temperature.

\end{enumerate}

This study provides a detailed examination of how the initial effective temperature of plasma species affects the energy transfer at the kinetic scales using a scale filtering approach. In addition, it sheds light on the role of normal and shear deformation,\textit{ i.e.}, components that together capture the incompressible effects of the plasma flow on the energy transfer. Future work will be to apply these results to different plasma environments, such as the magnetosheath and solar wind, and characterize the contribution of individual plasma species to the net energy 
transfer occurring at these scales.

Finally, it remains uncertain whether the conclusions drawn in this study apply to more comprehensive and broadly representative plasma environments. One immediate 
avenue of future work is the extension of these findings to a wider range of temperature ratios. Another important factor is the role of dimensionality; given the significant differences in plasma behavior across various models of three-dimensional turbulence, future work should use 3D simulations to study 
whether the present conclusions will hold universally across different parameter regimes. 

\begin{acknowledgments}
This research used resources of the National Energy Research Scientific Computing Center (NERSC), a U.S. Department of Energy Office of Science User Facility located at Lawrence Berkeley National Laboratory, operated under Contract DEAC02‐05CH11231 using NERSC Award FES‐ERCAP0027083, which was awarded to Paul Cassak. We also acknowledge the high-performance computing support from Cheyenne and Derecho provided by NCAR's Computational and Information Systems Laboratory, sponsored by the NSF.

\end{acknowledgments}

\section*{Data Availability Statement}
The data that support the findings of this study are openly available in Zenodo at \\
\href{https://doi.org/10.5281/zenodo.17333341}{https://doi.org/10.5281/zenodo.17333341}.

\bibliography{apssamp}

\end{document}